\documentclass[aps,prc,twocolumn,floatfix,showpacs,preprintnumbers,amsmath,amssymb,nofootinbib,groupedaddress]{revtex4-1}
\usepackage{color}
\usepackage{graphicx}
\usepackage{dcolumn}    % Align table columns on decimal point
\usepackage{multirow}
\usepackage{bm}         % bold math
\usepackage{sidecap}
\usepackage{hyperref}   % use for hypertext links, including those to external documents and URLs
\usepackage{xcolor}
\usepackage{amsmath}
\usepackage{hyperref} 
\definecolor{dark-red}{rgb}{0.,0.,0}
\definecolor{dark-blue}{rgb}{0.,0.,1}
\definecolor{medium-blue}{rgb}{0,0,1}
\hypersetup{
    colorlinks, linkcolor={dark-red},
    citecolor={dark-blue}, urlcolor={medium-blue}
}

\begin{document}
\title{A model for Ni-63 source for betavoltaic application}

\author{Abderrahmane Belghachi\textsuperscript{1}}
\email{abelghachi@yahoo.fr}
\author{Kutsal Bozkurt\textsuperscript{2}}
\author{Hassane Moughli\textsuperscript{1}}
\author{Orhan {\"{O}}zdemir\textsuperscript{2}}
\author{Benameur Amiri\textsuperscript{1}}
\author{Abdelkrim Talhi\textsuperscript{1}}

\affiliation{\textsuperscript{1}Laboratory of semiconductor devices physics (LPDS), University of Bechar, Bechar 08000, Algeria\\
             \textsuperscript{2}Yildiz Technical University, Davutpasa Campus, Science and Arts Faculty, Physics Department, 34220 Esenler Istanbul, Turkey 		
                              }

\date{\today} 

\begin{abstract}
A mathematical model of Ni-63 source for betavoltaic batteries is presented, based on Monte Carlo calculation. Trajectories of beta particles are simulated in Ni-63 source until their escape or total energy dissipation. Analysis of the effect of physical and technological factors on the performance of a source is carried out. Special attention is given to self-absorption and substrate backscattering because of their impact on power emission. Addition of a protective layer diminishes the source emission because of further absorption. The model has been tested successfully for Ni-63/GaN structure. 
\end{abstract}

\maketitle
\section{Introduction}
Conversion of nuclear decay energy into electrical power attracted significance attention since the early 1900s \cite{1, 2}. During last decades betavoltaic nuclear batteries have been the focus of intensive research work \cite{3,4,5,6,7,8}. Nuclear batteries are best candidates for a number of applications where long-life power sources or low energy consumption are required such as; space applications, pacemakers, microsystems, remote-sensors, etc. A betavoltaic cell consists mainly of a beta particles radioactive source and a semiconducting material with a pn, pin or Schottky junction. The fundamental operating principle of a betavoltaic device is the interaction of beta particles with matter releasing a substantial number of electron-hole pairs. The role of pn junction is the establishment of a build-in electric field insuring the separation of generated free carriers, therefore creating usable electric output power. Several radioisotope substances have been investigated as betavoltaic sources, for instance; H-3, S-35, Ni-63, Kr-85, Y-90, Pm-147, etc.  Among these sources, Ni-63 is the most promising choice because of its desirable qualities. Besides, of being pure beta source, Ni-63 has a long half-life (about 100 years), produces low energy beta particles, so minimising radiation damage to semiconductor converter, and can be stopped within few micrometres traveling in solids. As for its abundance, Ni-63 is important in the classification of radioactive waste from nuclear power plants \cite{9}. 
In a betavoltaic device, the radioactive source plays a very important role in the determination of the structure performance.  
Numerous work has been dedicated to betavoltaic batteries but little has been devoted to studying radioactive source component \cite{10}. The present work edifies the characteristics of beta source by investigating the correlation between physical and technological parameters to source output. The used quantity of radionuclide substance in a source has to be well defined in order to produce maximum power activity. Source thickness (amount of Ni-63) and geometrical form of the emitting surface significantly control the amount power to be delivered to the active semiconducting region. To model a radioactive source all physical phenomena occurring within the structure has to be accounted for. Two major factors affect source output; these are source self-absorption and substrate backscattering. No matter how thin is the active layer, absorption is inevitable, it results in reduction of emitted beta particles energy and number, this is self-absorption. A radioactive source is deposited always on a material that is called source backing material or substrate. This layer is generally a thin film, but no matter how thin, it may backscatter beta particles traveling away from the emitting surface. The suggested source model is obtained as a result of a Monte Carlo technique and assuming an attenuation absorption law. The model predicts Ni-63 source output parameters namely: apparent activity, emitted energy spectra and power density. Similar approach can be implemented to model different radioactive source and with different geometry. The model has been tested successfully for Ni-63/GaN structure, giving results matching well with previously simulated results  \cite{11}.

\section{Model}	

Monte Carlo technique have been widely used to simulate stochastic transport phenomena for the last few decades. Particularly with the development of high-level computing skills, reduction of computational times and increase of storage capacity. In order to get more insight into the transport properties of beta particles in matter a simplified MC approach was used. The developed Monte Carlo code simulates the interaction of generated electrons in a radioisotope substance with a solid material using the single scattering approximation. This model is reported to be an accurate representation of electron interaction and is capable of giving excellent results \cite{12}. For any standard Monte Carlo program, two phenomena must be modelled: elastic collision and energy loss.
When beta particles (electron) travels in a solid it could interact elastically with positively charged atomic nucleus or with atomic electrons in which case the electron will be deflected from its initial direction by Coulomb forces while its energy will remain unaltered. Alternatively, electron could interact inelastically with atoms by removing inner-shell electrons from orbit or with valence electrons to produce secondary electrons. In semiconducting materials, this latter scattering mechanism leads to the generation of electron-hole pairs. During their travelling in a solid, electrons are subject to a succession of scatterings, either elastically or inelastically. This process will continue until either the electron gives up all of its kinetic energy to the solid and comes to thermal equilibrium with it or until it manages to escape from the solid across its limiting surface.

The trajectory of each electron in a solid comprises of a series of random straight paths, their lengths and directions are determined essentially by scattering mechanisms probability. In the single scattering approximation, only elastic scatterings are implied in the determination of the path and direction taken by any given electron. Electron trajectory is a function of the elastic mean free path $\lambda_{el}$, which depend only upon the scattering rate. This rate is related to the total scattering cross section via
\begin{equation}
\lambda^{-1}_{el}=\rho N_{\alpha}\sum^{n}_{i=1}\frac{C_{i}\sigma_{el,i}}{A_{i}}.
\end{equation}             
Here, $N_\alpha$ is Avogadro's number, $\rho$ is the material density, $A_i$ is the atomic weight of element $i$, $C_i $ is the mass fraction of element $i$, $\sigma_{el,i}$ is the total cross-section for element $i$ and $n$ represents the number of elements forming the material. The free path (distance between two scatterings) is determined by the electron mean free path and a random number $r$ in a range $[0,1]$
\begin{equation}
l=-\lambda_{el}\log (r).
\end{equation}      
The elastic Mott total cross-section for each element is computed using the model given by \cite{13}, 
\begin{equation}
\sigma^{T}_{M}=5.21\times 10^{-21}\frac{Z^2}{E^2} \frac{4\pi \lambda \Big[1-e^{-\beta \sqrt{E}}\Big]}{\alpha(1+\alpha)}\Bigg [\frac{E+511}{e+1022} \Bigg]^2 ,
\end{equation}
where $\alpha$ is screening parameter given by
\begin{equation}
\alpha=3.4\times10^{-3}\frac{Z^{2/3}}{E} ,
\end{equation}    
$Z$ is the atomic number and E is the electron kinetic energy. The values of $\lambda$ and $\beta$ for each $Z$ are extracted from Ref. \cite{13} (table 1). In the case of polyatomic material, the element responsible of the scattering has to be determined. To achieve this we draw a random number $r$ and compare $r\times \sum^{n}_{i=1} F_{i}\sigma_{i}$ to the cumulative scattering mechanisms rate for $j$ varying from $1$ to $n$,
\begin{equation}
\sum^{j-1}_{i=1} F_{i}\sigma_{i}< r \times \sum^{n}_{i=1} F_{i}\sigma_{i} \leq \sum^{j}_{i=1} F_{i}\sigma_{i} ,~j=2,.....n .
\end{equation}     
Here, $F_i$ is the atomic fraction of element $i$ and $\sigma_i$ is the total cross-section of element $i$. When this inequality is verified, the scattering mechanism $j$ is then selected. The polar angle of an individual elastic collision $\theta$ is determined with the value of the partial cross section of the element $i$. $\theta$ is obtained by solving
\begin{equation}
r_{1}=\frac{\int^{\theta}_{0}\frac{d\sigma}{d\omega}\sin\theta d\theta}{\int^{\pi}_{0}\frac{d\sigma}{d\omega}\sin\theta d\theta} .
\end{equation}    
With $r_1$ a random number uniformly distributed between 0 and 1, a solution is found for the case of Rutherford theory \cite{12}
\begin{equation}
\cos\theta=1-\frac{2\times\alpha \times r_1}{(1+\alpha-r_1)} ,
\end{equation}    
with $\alpha$ a screening parameter is given by Eq.(4). The azimuthal angle after the elastic collision $\phi$ is selected by another random number $r_2$, $\phi$ is uniformly distributed between $0$ and $2\pi$
\begin{equation}
\phi=2\pi\times r_2
\end{equation}     
once  $\theta$  and $\phi$  are determined, we calculate the new direction $a_x$,  $a_y$ and $a_z (cosines)$ as function of previous direction $a_{x0}$,  $a_{y0}$ and $a_{z0}$ using spherical coordinate system transformation,
\begin{eqnarray}
a_x&&=\!a_{x0}cos\theta+\frac{a_{z0}a_{x0}}{\sqrt{1-a_{z0}^2}}\sin\theta\sin\varphi-\frac{a_{y0}}{\sqrt{1-a_{z0}^2}}\sin\theta\sin\varphi  \nonumber \\
a_y&&=a_{y0}cos\theta+\frac{a_{z0}a_{y0}}{\sqrt{1-a_{z0}^2}}\sin\theta\cos\varphi-\frac{a_{x0}}{\sqrt{1-a_{z0}^2}}\sin\theta\sin\varphi    \nonumber \\
a_z&&=a_{z0}cos\theta-\sqrt{1-a_{z0}^2}\sin\theta\cos\varphi .
\end{eqnarray}
The electron energy is assumed to dissipate steadily along its path at a rate governed by the well-known continuous loss approximation described by Bethe relationship \cite{14}
\begin{equation}
\frac{dE}{ds}=\frac{-7.85\times 10^{-3}\rho}{E}\times\sum^{n}_{i=1}\frac{C_{i}Z_i}{A_{i}}ln\Bigg[1.116\Big(\frac{E+k_i J_i}{J_i}\Big)\Bigg] ,
\end{equation} 
where $Z_i$ is the atomic number of element $i$, $C_i$ is the mass fraction of element $i$ $(C_i=m_i/m_{tot} )$, $F_i$ is the atomic fraction of element $i$, $k=0.734\times Z^{0.037}$ \cite{15} and $n$ is the number of elements in the region (it is the atomic $\%$ of element $i$ in mixture) with 
\begin{equation}
    J=
    \begin{cases}
      11.5\times Z & \text Z<13 \\
      9.76+58.5\times Z^{-0.19} & \text Z\geq13 .
    \end{cases}
  \end{equation}

Unlike most available Monte Carlo simulators designed to simulate monoenergetic electron beam of a finite section interaction with solids, our program deals with the interaction of a flux electron with an energy spectra impinging a solid through its surface. The program simulates also radioisotope substances where electrons are spontaneously generated within the material bulk. This program is specifically designed for betavoltaic devices, which consist mainly of a radioactive source layer together with a thin film semiconductor. 

To simulate a betavoltaic battery we chose a common structure, which consists of a radioactive layer laid onto a semiconductor p-n junction rectangular bloc of thickness $t$ and a unit surface $(1cm\times1cm)$.  The structure is partitioned into adjacent unit square cells with $1\mu m\times1\mu m$ surface and a thickness $t$ that are assumed identical. This unit cell is sliced into a stack of wafers of $1\mu m\times1\mu m$ area and of thickness $\Delta z$, assuming that the total device thickness is very small compared to its surface, therefore perimeter effect could ignored. 

The most important parameter for betavoltaic cell is the spatial distribution of energy deposition in semiconductor that is responsible for electron-hole generation. Spatial distribution of energy deposition in the structure is computed as a function of thickness for unit surface. Only electrons generated inside the source region of the cell are followed even across lateral cell boundaries (sidewalls), this ensures continuity between the simulated cell and its neighbour cells.  

Lateral coordinates $(x, y)$ and normal direction $z$ coordinate describe electron position. Only the $z$ direction is discretised $(z_i=i\times\Delta z)$, normal to the device stacking layers, where deposited energy needs to be known. Energy dissipation rate will change every time electron cross from one region to another. 

Deposited energy in a wafer $i$ is computed as the sum of the dissipated electrons energy ($\sum_{j}\Delta E^{j}_{i}$ where $j$ represents electron number) while crossing the wafer $\Delta E^{j}_{i}=\frac{dE}{ds}\lambda_{ij}$  where $\lambda_{i,j}$  is part of free path of electron $j$ inside the wafer $i$.

The first part of this work we aim to simulate, using an in-house code, the output of Ni-63 source and analyse some factors that affect its behaviour, essentially emitted power density and energy spectra. Most of the available simulators were developed for monoenergetic electron beam, whereas in a real beta source the spectrum of emitted particle energy is rather complex. In our simulation we adopted the beta spectra data obtained by \cite{16}, which reproduced well the measured beta spectra of Ni-63. The simulated source consisted of a thin Ni-63 active layer, with a varying thickness, deposited onto Ni metallic foil as a substrate and a protective Ni layer was optionally added. 

In the second part, we simulated a Ni-63 source deposited onto a 4 $\mu m$ thick GaN thin film to investigate the effect of source thickness on the deposited energy. In the simulation we used $10^{6}$ electrons. A full energy spectrum of Ni-63 is used (extracted from BetaShape Analytical version: 1.0 (24/06/2016)\cite{17}) and a specific activity of $2.1\times 10^{12}$ $Bq /g$. The simulated source is $1cm \times 1cm$ and a thickness ranging from 0.01 to 7 $\mu m$. Total history of each electron (position, direction and kinetic energy) is computed from its site of creation inside the radioactive Ni-63 region until it exits from the back of the cell (backscattering is included) or comes to thermal equilibrium (that is when its energy decreases to below 0.05 $keV$). 
\begin{figure}
\centering
   \includegraphics[width=0.5\textwidth]{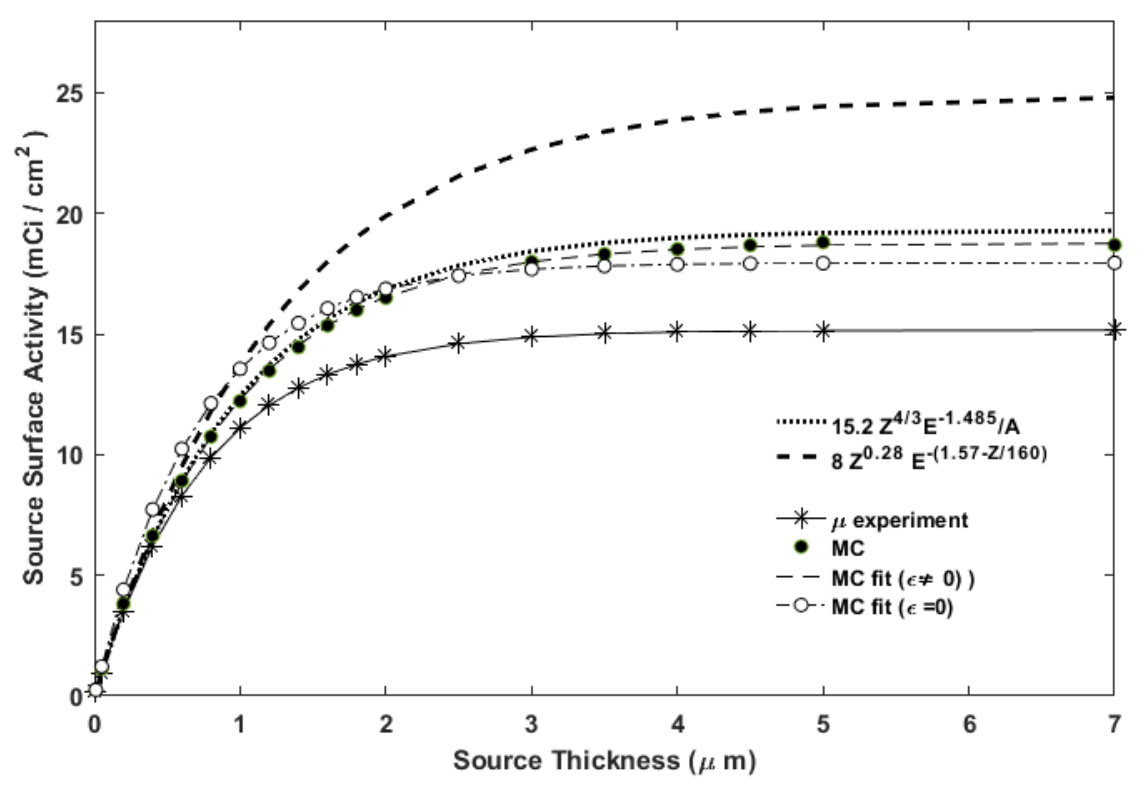}
  \caption{Source surface activity of Ni-63 versus source thickness for different $\frac{\mu}{\rho}$ expressions with our Monte Carlo simulation result (both fitting curves are shown with $\varepsilon = 0$ and $\varepsilon \neq 0 $).}
\end{figure}
\begin{figure}
  \centering
  \includegraphics[width=0.5 \textwidth]{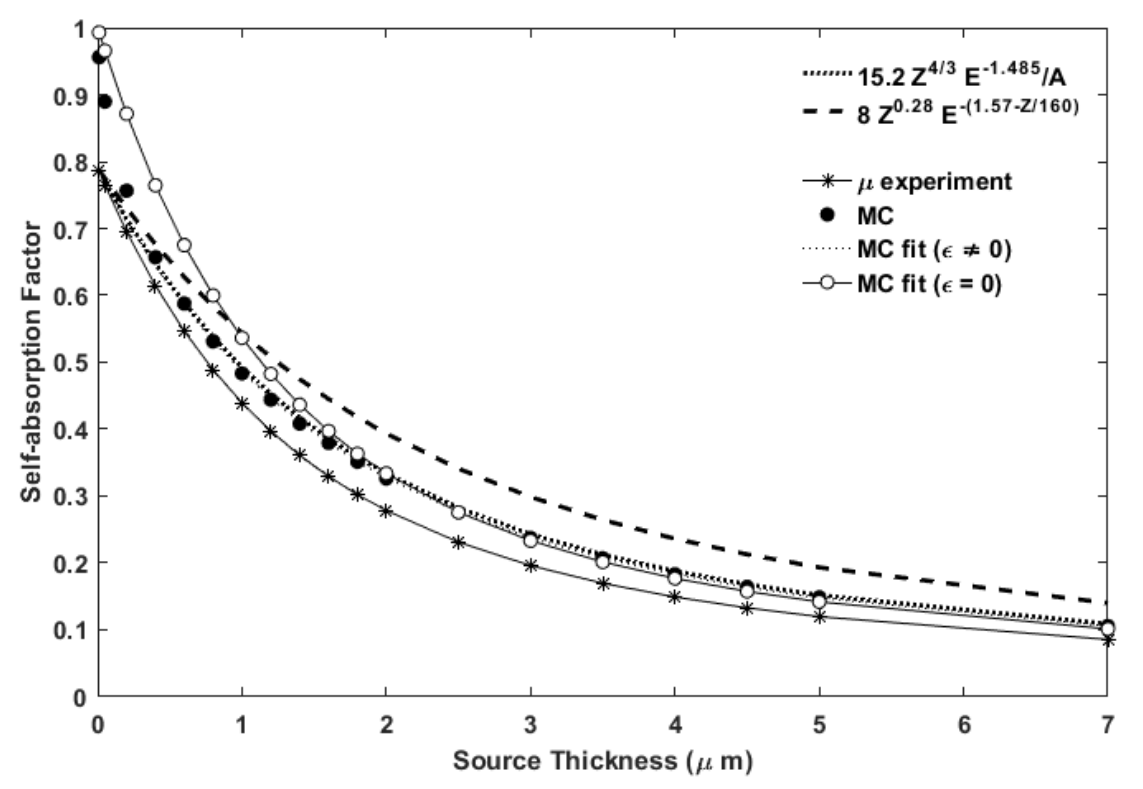}
  \caption{ Self-absorption factor curves for Ni-63 versus source thickness for different $\frac{\mu}{\rho}$ expressions together with our Monte Carlo simulation result (both fitting curves are shown with $\varepsilon = 0$ and $\varepsilon \neq 0 $).}
\end{figure}
\begin{figure}
\centering
\includegraphics[width=0.5\textwidth]{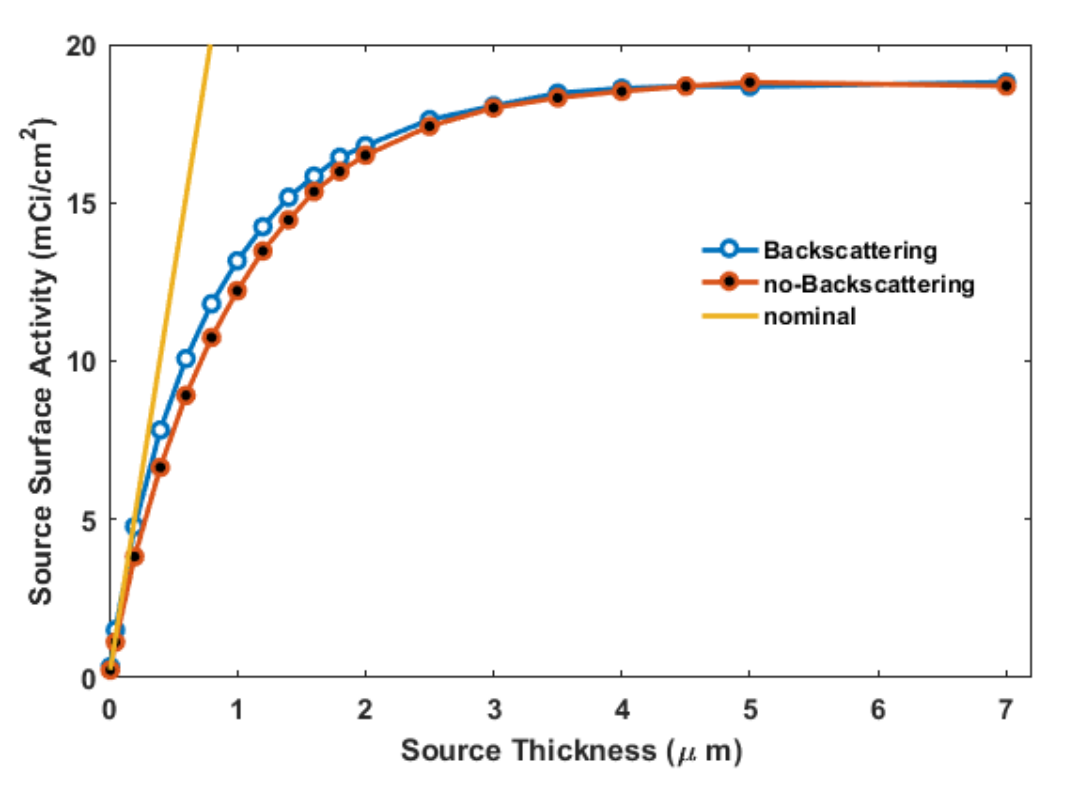}
  \caption{Source surface activity vs. thickness with and without backscattering effect, obtained using Monte Carlo simulation, together with nominal activity.}
\end{figure}

\section{Results and Discussion}
	
\subsection{Source self-absorption effect}	
To investigate source self-absorption, we simulate a rectangular slab of bare Ni-63 radioisotope material, without substrate nor protective layer, with different thicknesses.  
Due to the isotropic nature of radioactive substance, the external surface of a uniform mass will radiate equally in all directions. Let us consider a rectangular slab source with a thickness t and with a top and bottom surfaces $S$, so as its volume is $S\times t$ and a total external surface of $2S+4t \sqrt{S} $ ($S$ is assumed square shape); if its specific activity is $A_s$ ($Bq /g$), with the absence of absorption, nominal activity density can be expressed as  
\begin{equation}
A_{n}=\frac{A_s \rho S t}{2S+4t \sqrt{S}}=A_s \rho t\frac{1}{2+\frac{4t}{ \sqrt{S}}} .
\end{equation}
So
\begin{equation}
A_{n}=A_s \rho t \alpha_0,
\end{equation}                                          
where $ \alpha_0$ represent a form (geometric) factor in absence of self-absorption, then: 
\begin{equation}
 \alpha_0=\frac{1}{2+\frac{4t}{ \sqrt{S}}}
\end{equation}
$ \alpha_0$ is a factor always less or equal to 0.5. 
\begin{figure}
  \centering
  \includegraphics[width=0.5 \textwidth]{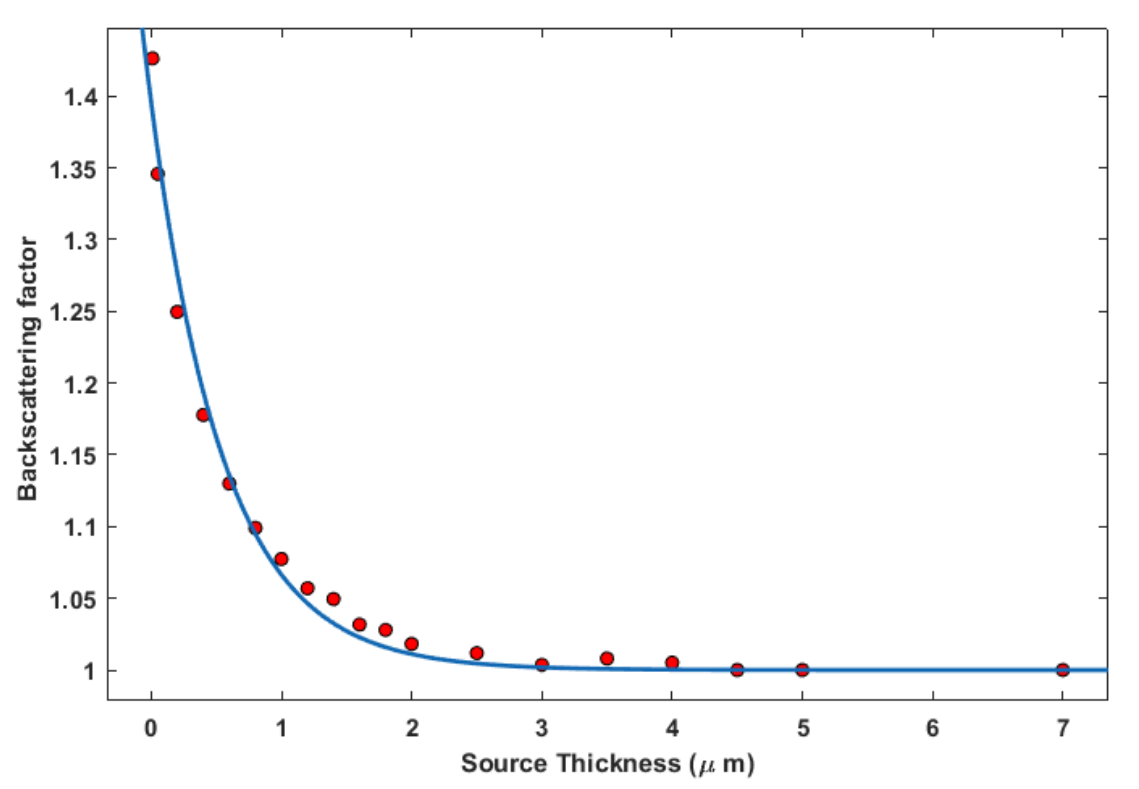}
  \caption{ Monte Carlo calculation of the backscattering factor against Ni-63 source thickness with the best fit using equation (22) $a$=0.395, and  $b$=1.78 $\times$ 10$^4$ $cm^{-1}$.}
\end{figure}	
 \begin{figure*}
  \begin{center}
\includegraphics[width=0.9\linewidth,clip=true]{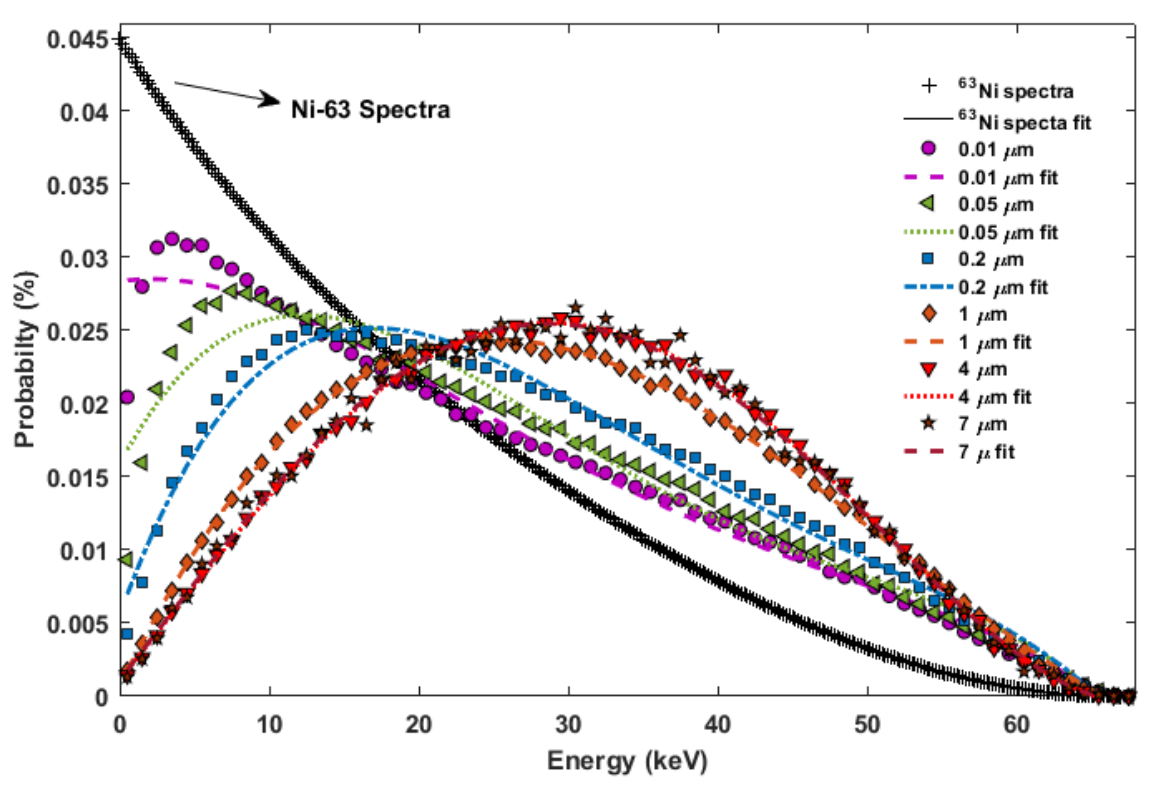}
  \end{center}
  \caption{Effect of source thickness on beta energy spectrum at the source emitting surface, solid line represents energy spectrum of Ni-63 radioisotope as calculated by X. Mougeot \cite{22}.}
\end{figure*}
	
In the case of a disc slab source of radius r (cylindrical shape), the form factor is then:
\begin{equation}
 \alpha_0=\frac{1}{2+\frac{2t}{r}}.
\end{equation}
For thin film sources, where thickness $t$ is very small compared to $S$, $\alpha_0$ could be approximated by $\alpha_0 \approx 0.5$ then $A_n = 0.5 \times A_s \rho t$.     

The increase of beta source thickness will increase the amount of emitted electrons, whereas source self-absorption progressively overcomes.

Source self-absorption will result in a reduction of kinetic energy of emitted electrons in addition to altering their number. To estimate self-absorption effect we assumed that the source was a uniform deposit of radioisotope Ni-63 emitting beta particles of a specific activity $A_s$ and the emitted particles followed a known attenuation law \cite{18}. Apparent activity of a source without backscattering is related to source thickness according to the following law
  \begin{equation}
 A(t)=\frac{A_s \rho\alpha}{\mu} (1-e^{-\mu t}),
\end{equation}   
where $\alpha$ represents the new form factor taking into account source self-absorption, we propose an expression for this geometrical factor as follows     
   \begin{equation}
\alpha=\alpha_0+\varepsilon ,
\end{equation}   
then
 \begin{equation}
 A(t)=\frac{A_s \rho(\alpha_0+\varepsilon)}{\mu} (1-e^{-\mu t}).
\end{equation}   
In these equations  $\varepsilon$  is assumed constant over the simulated thickness, its value is obtained for the best fit to our calculated activity $A(t)$ curve and $\mu$ is the linear attenuation coefficient of Ni-63.

Values of attenuation coefficients are determined, either experimentally or using empirical formulas from literature. These Values are scattered in literature and cover a wide range. Discrepancy of attenuation coefficient in literature could be attributed to material quality; besides geometrical form of structures do have some effect on it \cite{19}.  

With Monte Carlo simulation, we computed surface activity of Ni-63 source without backscattering from the backing Ni metal and without protective layer. The variation of source activity against thickness curve when fitted to a common attenuation law (Eq. (18) with  $\varepsilon =0$) allow the extraction of linear attenuation coefficient (absorption coefficient) $\mu$, for this case we found $\mu =$ 14080 $cm^{-1}$, this value appears to be high compared to most reported data. The closest value is found in Ref. \cite{19} where they reported an experimental average of $\mu$ (13183.84 $cm^{-1}$). Using Eq. (18) ($\varepsilon\neq 0$) the best fit to our results gives $\mu = $10650 $cm^{-1}$ which compares well with data obtained from most previously reported attenuation coefficient expressions, as summarised in table 1. In figure 1 we plotted surface activity density obtained by our simulation with both fittings ($\varepsilon =0$  and $\varepsilon\neq 0$) together with a recap of obtained results using Eq. (18) for different values from several $\mu$ formulas (see table 1). The value of $\varepsilon$ corresponding to the best fit was  $\varepsilon = - 0.1046$, this value is valid exclusively for our simulated source (i.e. $1cm\times1cm$ profile). 

Self-absorption in a source is characterised by a factor, $f_{sa}$ defined as a ratio of the number of particles leaving source with self-absorption (apparent activity) to the number of particles leaving source without self-absorption (nominal activity) so:
\begin{equation}
 f_{sa}=\frac{\alpha}{\mu t \alpha_0} (1-e^{-\mu t}),
\end{equation}             
 \begin{equation}
 f_{sa}=\frac{(1+2\varepsilon)}{\mu t} (1-e^{-\mu t}).
\end{equation}           
Figure 2 represents the variation of  $f_{sa}$ against source thickness. The parameter $\varepsilon$  obtained from activity versus source thickness cure is used to evaluate $f_{sa}$ using two available $\mu$ expressions and an experimental value from references using Eq.(19).         
  \begin{table}
\centering
\caption{Summary of   $\frac{\mu}{\rho}$  extracted from computed curves and values from literature.}
\tabcolsep=0.4cm
\def\arraystretch{1.5}
\label{t3}
\begin{tabular}{lllll}
\hline\hline
Expression of                                 $\frac{\mu}{\rho}$    & $\mu$  $(cm^{-1})$  \\ \hline
15.2 $Z^{4/3}$ $\frac{E^{-1.485}}{A}$   \cite{20}     &10358.32 \\
8 $Z^{0.28}$   $E^{-(1.57-\frac{Z}{160})}$ \cite{21}           & 8031.75 \\ 
Experimental \cite{19}                                      & 13183.84   \\ 
Monte Carlo calculation ~~~~~$\varepsilon$ = - 0.1046       & 10650                        \\ 
~~~~~~~~~~~~~~~~~~~~~~~~~~~~~~~~~~~~~$\varepsilon$ = 0        & 14080  \\\hline\hline
\end{tabular}
\end{table}
          
 \subsection{Substrate back-scattering effect}
 	           
The substrate or source backing material may backscatter beta particles, therefore contributing to the flux of emitted particles from the source front. To investigate this effect we introduced in our Monte Carlo program the backing material scattering, which is a layer of metallic Ni. The back-scattered beta particles from the substrate or the backing material are reintroduced into the active layer and allowed to diffuse accordingly. 

In figure 3 we represented source surface activity with and without backscattering effect together with nominal activity. Backscattering effect is observable at low thicknesses where source activity is almost linearly proportional to thickness. As saturation is attained backscattering has no effect and source self-absorption dominates. 

To separate backscattering effect from self-absorption effect it is necessary to examine the backscattering factor $f_{bs}$  which is defined as the rate of number of particles leaving the source with source backing to the number of particles leaving without source backing. 

In the case of total reflection and negligible thickness $f_{bs}$  should be equal to 2, for thick sources where self-absorption dominates $f_{bs}$  tends to 1, therefore $1\leq f_{bs}\leq 2$. So the apparent source activity is written as
 \begin{equation}
 A(t)= f_{bs}\frac{A_{s}\rho(\alpha_0+\varepsilon)}{\mu}(1-e^{-\mu t}).
\end{equation}            
 
For the investigated source, the variation of backscattering factor versus thickness is represented in figure 4. As expected $f_{bs}$ decreases sharply to saturate around unity. In this figure, we plotted a function $f_{bs}(t)$ that best fits Monte Carlo obtained results. This function has two fitting parameters $a$ and $b$,             
where
 \begin{equation}
f_{bs}(t)=1+a\times e^{-bt}.
\end{equation}             
Our best fit to the computed results yields: $a$=0.395 and  $b$=1.78 $\times$ 10$^4$ $cm^{-1}$.
 \begin{figure}
  \centering
  \includegraphics[width=0.5 \textwidth]{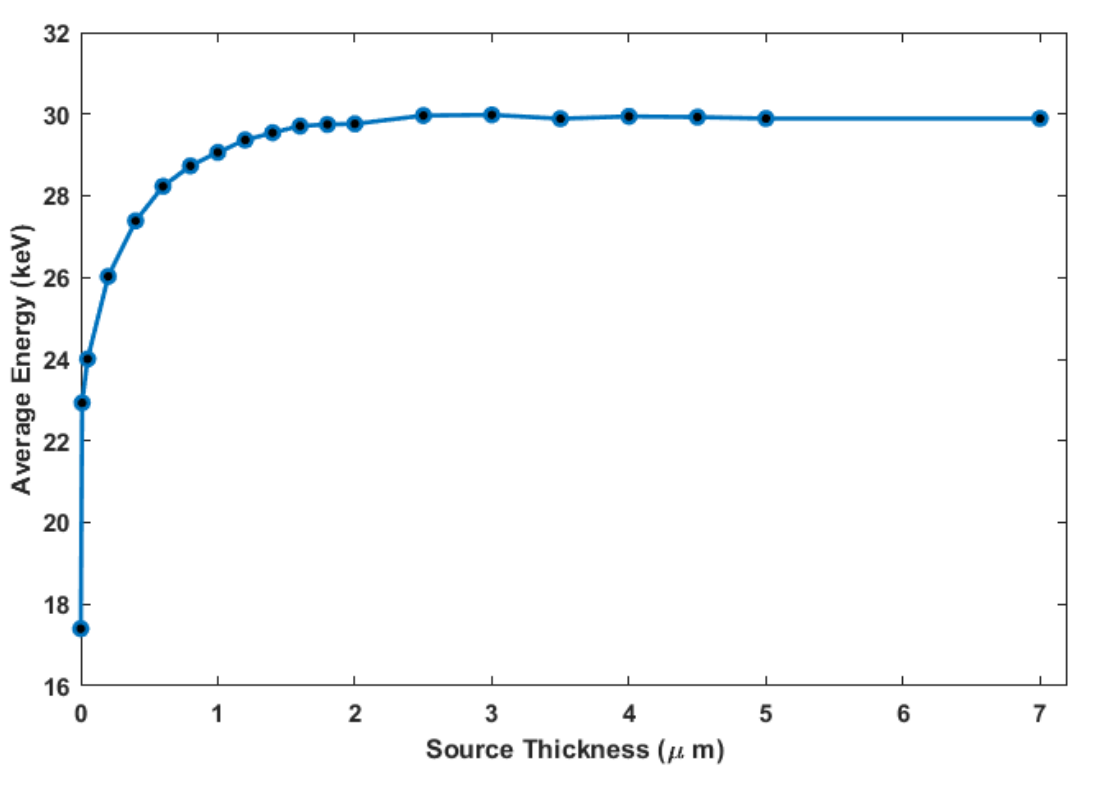}
  \caption{Effect of source thickness on the surface emitted beta particles average energy.}
\end{figure}
	
  \subsection{Energy Spectrum}
 Each radioisotope material have a continuous energy probability distribution known as beta energy spectrum. The beta spectra emitted by Ni-63 radioisotope recently measured down to very low energies, are well reproduced by the calculations of X. Mougeot et al. \cite{16, 22}. It is characterised by emission of electron ($\beta$-particle) having average energy 17.4 $keV$ and a maximum energy of 66.9 $keV$ diffusing in random directions. 
 
Energy probability distribution has been described by an expression of the following form \cite{23}
  \begin{equation}
W(E)\propto (E^{2}+2E m_{e}c^{2})^{1/2}(Q-E)^{2}(E+m_{e}c^{2}),
\end{equation}     
where $E$ is the electron kinetic energy, $m_e$ is the electron mass, $Q$ is the decay energy, and $c$ is the speed of light. 

In our study we proposed an approximation for the probability of beta particles escaping from a Ni-63 source, which can be expressed in the following formula 
 \begin{equation}
W(E)=a_{0}+a_{1}E+a_{2}E^{2}+a_{3}E^{3}+a_{4}E^{4},
\end{equation}   
where $a_{i}$ coefficients depending on source thickness and are determined from $W(E)$ curves. 

In figure 5, we represented energy beta spectrum at the surface of Ni-63 source for different thicknesses together with a spectrum from the Ni-63 decay.  It shows that as the source thickness is increased the maximum of energy distribution is shifted towards higher energies. This is due to source self-absorption; the kinetic energy of emitted electrons is diminished in addition to a reduction of their number.
  
$a_{i}$   coefficients are extracted from calculated curves using best fit to the obtained spectrum using a 4th degree polynomial curve (Eq. (24)). These values saturates as the thickness of source reaches about 4 $\mu m$ where spectra curves coincides, table 2 summarises $a_{i}$  values.  At saturation Ni-63 emitted beta particles spectrum at the source surface, independently of source thickness, could be described by the following equation
\begin{eqnarray}
W(E)&&=\!0.0011+0.0013\times E+4.6\times 10^{-6}\times E^{2} \nonumber \\
&&-1\times 10^{-6}\times E^{3}+1\times 10^{-8}\times E^{4}.
\end{eqnarray}

\begin{table}
\centering
\caption{Summary of  $a_i$   values for different source thicknesses extracted from best fit to calculated curves to equation (24).}
\label{t2}
\tabcolsep=0.1cm
\def\arraystretch{1.5}
\begin{tabular}{llllllllll}
\hline\hline
t ($\mu m$) & ~~$a_0$&~~$a_1$ &~~~$a_2$ &~~~$a_3$ &~~~$a_4$   \\ \hline
0	 	 &0.04476        & -0.00154     & 2.53e-05       & -3.31e-07      & 2.22e-09       \\

0.01		 &0.02835        & 0.00015      &-3.82e-05      & 8.25e-07        & -5.78e-09       \\

0.05		&0.01590        & 0.00191      & -0.00011       & 2.07e-06        & -1.32e-08        \\ 

0.2		&0.00558        & 0.00280      & 0.00013       & 2.14e-06        & -1.27e-08          \\ 

1		 &0.00068       & 0.00209      & -5.40e-05       & 3.14e-07      & 3.09e-10           \\ 

4		 & 0.00110       & 0.00127      &7.86e-06       &-1.08e-06       & 1.00e-08            \\ 

7		 &0.00110        & 0.00132     & 4.63e-06       &-1.01e-06       & 9.60e-08           \\ \hline\hline

\end{tabular}
\end{table}
  \subsection{Average energy}
  As the energy spectrum shifted towards higher energies, the average energy increased up to a thickness of about 2 $\mu m$ where it saturates around 30 $keV$, as shown in figure 6. The average energy of emitted beta particles at a source of thickness $t$ $(E_{av} (t))$ is computed using
 \begin{figure}
  \centering
  \includegraphics[width=0.5 \textwidth]{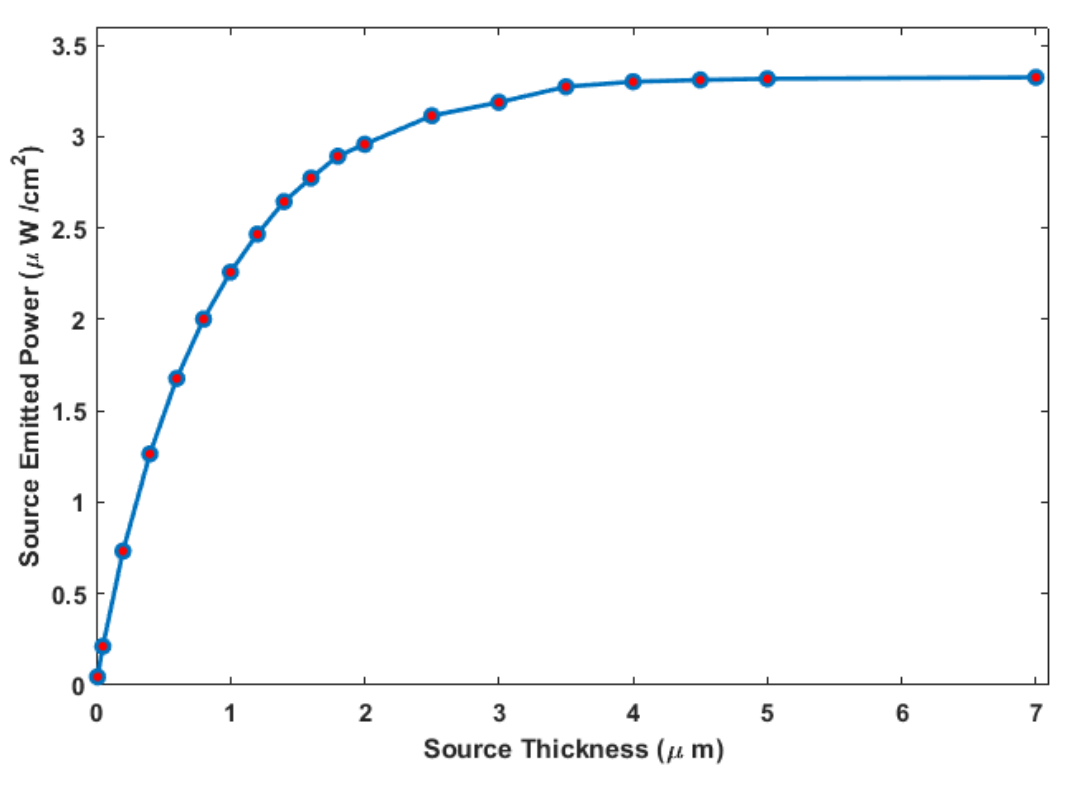}
  \caption{ Effect of source thickness on the source emitted power density.}
\end{figure}	 
  \begin{equation}
E_{av}(t)=\int_{0}^{E_{max}} W(E)EdE,
\end{equation}   
where $W(E)$ is the occupation energy probability and $E_{max}$ is the maximum energy of beta spectrum ($E_{max}$ = 66.9 $keV$).  
 \subsection{Emitted power density}  
 Source emitted power is very important for betavoltaic application, it provides the incident power density available for energy conversion. The total emitted power density from a source with a thickness t is computed using the following expression \cite{11}
\begin{equation}
P(t)=A(t) E_{av}(t)q,
\end{equation}  
where  $A(t)$ is the density of source surface activity $(Bq/cm^{2})$ and $q$ electron elementary charge. 
\begin{figure}
  \centering
  \includegraphics[width=0.5 \textwidth]{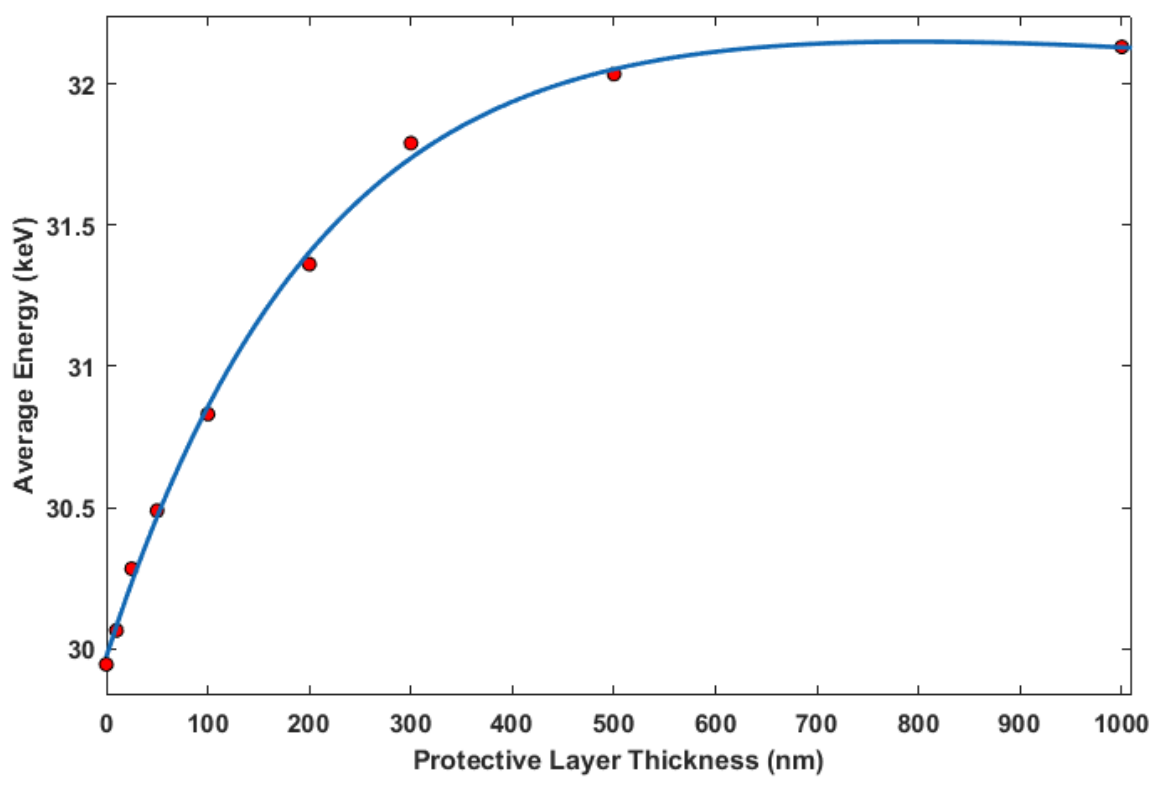}
  \caption{ Effect of the protective layer thickness on the beta average energy for an Ni-63 source of a 4 $\mu m$ thickness.}
\end{figure}	
In figure 7 we plotted emitted  power density at the source surface against its thickness. The curve shows a linear increase at low thicknesses then saturates at about 3.28 $\mu W/cm^2$ for a thickness above 3.5$\mu m$. Similarly, Ref. \cite{10} found 3.24 $\mu W/cm^2$ as apparent power density at saturation. The results compare well with reported data of Ref. \cite{11}, where they found 2.85 $\mu W/cm^2$ for 1.5 $\mu m$ source thickness with a specific activity of $2.2\times10^{12}$ $Bq/g$ (our results are 2.75 $\mu W/cm^2$ for the same thickness taking a specific activity of $2.1\times10^{12}$ $Bq/g$). 
  
\subsection{Effect of protective layer on the propriety of Ni-63 source}    
In the next simulation, a protective layer is added to the source so the emitting area will be the external protective layer surface. We investigated the effect of a metallic Ni layer as a protective layer onto a Ni-63 source of a 4 $\mu m$ thickness. The protective layer thickness is varied from 0 to 1000 $nm$. 

Figure 8 shows the variation of the average energy versus the protective layer thickness. We remark a significant increase of the average energy before saturation, meaning that the spectrum is further shifted towards higher energies. This is due to additional protective layer absorption effect. This effect is confirmed in figure 9, which represents the emitted power of the source versus Ni layer thickness. The emitted power density is seriously reduced if a thick protective layer is used. 

\begin{figure}
  \centering
  \includegraphics[width=0.5 \textwidth]{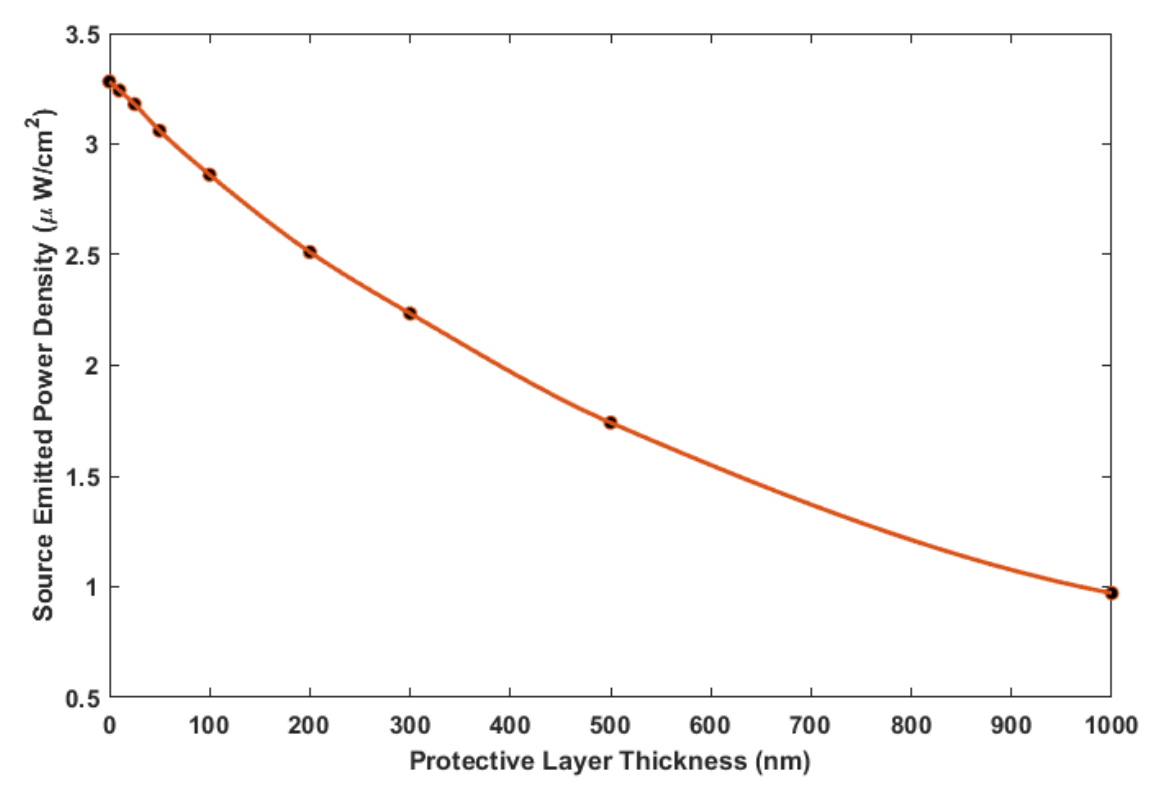}
  \caption{ Effect of the protective layer thickness on the emitted power of a Ni-63 source of a 4 $\mu m$ thickness.}
\end{figure}	
 \begin{figure}
  \centering
  \includegraphics[width=0.5\textwidth]{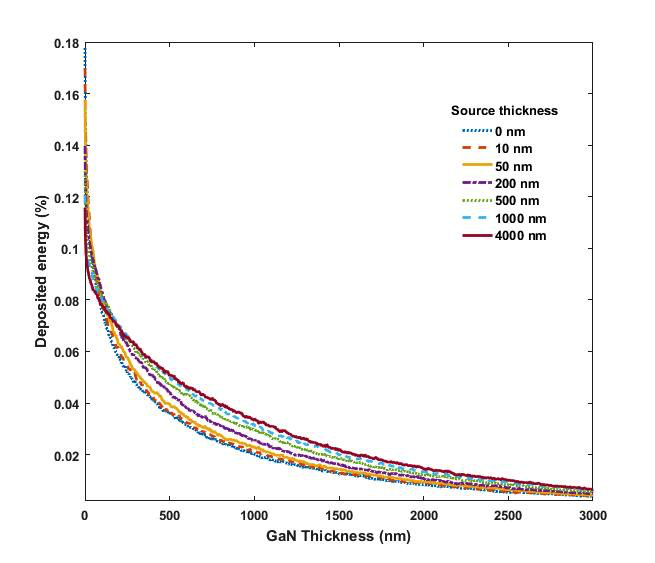}
  \caption{ GaN beta particle absorption profiles for various 63-Ni thicknesses.}
\end{figure}

\subsection{Deposited energy in GaN thin film}  
In this section, we investigate the effect of source thickness (from 0 to 4000 nm) on the absorbed energy in a GaN betavoltaic structure. To simulate Ni-63 source the energy spectrum obtained in section 2.3 is the starting point. Random initial energies for a number $N$ of emitted electrons from source are obtained using a cumulative distribution function CDF (integral of energy probability distribution). The results are plotted in figure 10, the curves show two regions: a)- on the vicinity of source/GaN interface (below 150 nm), as the source thickness is decreased the more percentage energy is transmitted to GaN material, b)- far from the interface (greater than 150 nm) the opposite behaviour is observed. This could be simply explained by the fact that as sources thickness is augmented self-absorption significantly increases. On the other hand, for thicker sources the average energy increases therefore more energy is pumped into GaN. The obtained results, figure 10, are very similar to those obtained by C. E. Munson et al. \cite{11} using a model, they previously developed \cite{7}.

\section{Conclusions}
This study presented a mathematical model of Ni-63 beta particles sources. In this model we demonstrated the crucial effect of self-absorption which limits power emission beyond 4$\mu m$ thickness. We observed energy spectra of emitted electrons shifting towards higher energies and convoyed higher average energies up to saturation around 4$\mu m$. The suggested model consists of a modified attenuation law for the source output activity (apparent) to take into account self-absorption. Back-scattering effect from source substrate (backing support) has a remarkable effect for very thin active layers, this is characterised by a factor $f_{bs}$. For ultra-thin Ni-63 layers this factor tends to 2, therefore doubling the source activity. In the case where a protective layer is deposited atop of the source this will reduce significantly source activity and if its thickness is increased it will degrade the source overall performance.The suggested model has been tested to determine deposition energy in GaN and gave results similar to those obtained previously. The proposed model reproduced well previously reported Ni-63 source outputs and can be applied, with few adjustments, to other kind of radioactive sources.     

\section{Acknowledgements}
This work was carried out in the framework of a joint project between the Laboratory of Physics of Semiconductors Devices (LPDS) of Béchar University, Algeria and the group of Betavoltaic Energy Converter of Physics Department, Yildiz Technical University Istanbul, Turkey.

% >>>>>>>>>>>>>>>>>>>>>>>>>>>>>>>>>>>>>>>>>>>>>>>>>>>>>>>>>>>>>>>>>>>>
% REFERENCES.

%
% >>>>>>>>>>>>>>>>>>>>>>>>>>>>>>>>>>>>>>>>>>>>>>>>>>>>>>>>>>>>>>>>>>>>
\end{document}